%% file: paper_accepted.tex
\documentclass[11pt]{aastex}
\usepackage{graphics}
\usepackage{natbib}
\usepackage{rotating}

\title{Accretion of Jupiter's Atmosphere from a Supernova-Contaminated Molecular Cloud\\ \ \\
Submitted to Icarus 14-Oct-2008\\ \ \\
Revised 24-Nov-2009\\
Accepted 9-Feb-2010 \\
}

\shorttitle{Jupiter's Atmosphere from Supernova}
\shortauthors{Throop \& Bally}

\begin{document}

\input htmac.tex

\author{Henry B. Throop}
\affil{Southwest Research Institute}
\affil{1050 Walnut St, Ste 300, Boulder, CO  80302}
\email{throop@boulder.swri.edu}

\author{John Bally}
\affil{Center for Astrophysics and Space Astronomy}
\affil{University of Colorado, Boulder}
\affil{UCB 389, Boulder, CO  80309-0389}

\begin{abstract}
If Jupiter and the Sun both formed directly from the same well-mixed proto-solar nebula, then their atmospheric
compositions should be similar.  However, direct sampling of Jupiter's troposphere indicates that it is enriched in
elements such as C, N, S, Ar, Kr, and Xe by 2--6$\times$ relative to the Sun \citep{wla08}.  Most existing models to
explain this enrichment require an extremely cold proto-solar nebula which allows these heavy elements to condense, and
cannot easily explain the observed variations between these species.  We find that Jupiter's atmospheric
composition may be explained if the Solar System's disk heterogeneously accretes small amounts of enriched material such
as supernova ejecta from the interstellar medium during Jupiter's formation.  Our results are similar to, but
substantially larger than, isotopic anomalies in terrestrial material that indicate the Solar System formed from
multiple distinct reservoirs of material simultaneously with one or more nearby supernovas \citep[\eg][]{tba07}.   Such
temporal and spatial heterogeneities could have been common at the time of the Solar System's formation, rather than the
cloud having a purely well-mixed `solar nebula' composition.


\end{abstract}

\section{Introduction}


The Solar System's composition reflects that of the initial cloud core from which it formed.  On small bodies such as
the terrestrial planets, subsequent thermal and chemical processes have altered the composition.  However, for the Solar
System's two largest bodies -- the Sun and Jupiter (mass 1~\mjup\ $\approx$ 0.001~\msol) -- the
atmospheric composition is relatively stable against change since the time of formation.  In particular, the relative
atmospheric abundances of gases carbon (C), nitrogen (N), sulfur (S), argon (Ar), krypton (Kr), and xenon (Xe) are
believed to be fixed at the time of formation 4.5 billion years ago, and not affected substantially by subsequent
evolution.  The noble gases are stable against loss due to Jupiter's high escape velocity, and stable against sinking by
condensation because they don't combine with other atomic species.  Elements C, N, and S combine into larger molecules
such as CH$_4$, NH$_3$, and H$_2$S, but are believed to remain well mixed in Jupiter's troposphere without settling to
the interior.

Jupiter's composition at 5--20 bars was directly measured by the Galileo descent probe's mass spectrometer in 1995.
Initial results \citep[][hereafter AMN03]{amn03} showed that Jupiter was enhanced by $\sim3\pm1\times$ relative to the
then-current solar abundances of \citet[][hereafter AG89]{ag89}.  More recent work has better determined the Solar
abundances, leading to revised Jupiter:Solar abundances.  Somewhat surprisingly, these new results actually increased
the scatter between the elements, to the broader range of $4\pm2\times$ \citep[][hereafter WLA08]{wla08}.  The
largest change occurred in the abundance of Jovian Ar, which more than doubled to $5.4\pm1.1\times$ solar
\citep[][hereafter GAS07]{gas07}.  A subsequent measurement of solar Ar reduces it again to $2.45\pm0.7\times$
\citep{lod08}(hereafter Lod08).  All enrichments in this paper are in terms of number density relative to hydrogen,
normalized to the Sun; \ie, $(n_i/n_H) / (n_i/n_H)_{\rm{sun}}$.   

\section{Previous models}

\label{sect:previous}

Several previous models have been proposed to address Jupiter's `metallicity problem.'  All are based on condensation of
the volatile elements into ices, which can then be concentrated relative to hydrogen.  The `amorphous ice model'
\citep{oe06,omn99} proposes that Jupiter's atmosphere was formed from a combination of solar nebula material and `solar
composition icy planetesimals' (SCIPs). The SCIPs were made of amorphous water ice in which the other volatile species
condensed; these SCIPs had the same composition as the Sun, except were depleted in H.  Jupiter then formed from a
linear combination of these two components: when SCIPs were introduced into Jupiter from the protoplanetary disk, they
were heated and released their volatiles into Jupiter's atmosphere.  (This two-component model is distinct from the
formation of Jupiter's core, which is a separate issue.)  The `clathrate hydrate model' \citep{hgt08,hgl04} uses a
similar approach, but proposes that the volatiles are trapped within clathrate hydrates, rather than amorphous water
ice.  In this model, species condense out as the nebula slowly cools to their clathration temperature.  The process can
work at slightly warmer temperatures than the amorphous ice model.  The condensation order of hydrates leads naturally
to the Jovian abundance.  Both models have the appeal of simplicity and directness.


These models share a common problem, that the nebula must be very cold for volatiles to condense.  The condensation
temperatures required are 20--35~K for the noble gases onto amorphous ice or into solids, and 35--55~K for the clathrate
hydrates \citep{igh03,hgl04}.  The clathrate models assume very cold nebula temperatures of $<$20~K at 5~AU at several
Myr \citep{hgl04}.  These temperatures are quite low compared with both observations and models of disks.  For instance,
\citet{wsw07} measures 20--50~K at 50~AU, and much warmer inward, and \citet{dgs03} measures 50~K at 100~AU.  Models by
\citet{cg97} predict 75~K at 5~AU and 55~K at 10~AU; work by \citet{dd05} assumes 90~K at 5~AU and 30~K at 50~AU; and
the model of \citep{jdh07} predicts 50~K at 100~AU.  Large uncertainties exist in both observational and theoretical
studies of disks, and it remains to be established that temperatures can indeed reach the low temperatures needed by
these models.  The low disk temperatures required are also dangerously close to the 20--50~K gas and dust temperatures
of the molecular cloud itself, which set an absolute lower limit for the disk temperatures \citep{jb06,jmm06,bll91}.
The molecular cloud temperatures are highest in the largest clouds and those with massive stars, which are the regions
most likely to be where the Solar System formed \citep[][]{hdh04}.  

Several models address the temperature problem by forming solids not at their current location, but further
out in the nebula (\eg, 10--50~AU) where temperatures were lower.  Radial migration subsequently brought them to 5~AU.
For instance, \citet{amb05} explores the possibility that Jupiter formed in its entirety at 10--15~AU via the clathrate
model, and then migrated inward.  \citet{gh06} use photo-evaporation and viscous migration of the entire nebula to
explain Jupiter's enrichment.  In this work, UV flux from internal and/or external stars heats and preferentially
removes hydrogen from the inner proto-planetary disk.  These planetesimals then migrate inward where they heat and go
into forming Jupiter.  This model is appealing, and consistent with the idea that the young solar system may have
experienced the effects of nearby massive stars \citep[\eg,][]{th03,tbe01}.  

Bringing material from much further out, where the nebula remains cold, is appealing but there is a strong dynamical
constraint on how much material from the coldest regions of the disk could be incorporated into the disk. 
\citet{ld97} showed that less than 0.5\% of comets leaving the Kuiper Belt eventually impact Jupiter.  To
acquire the amount of contamination seen here would require that the Kuiper Belt have a mass on the order 2500 Earth
masses.  Models of Neptune's outward radial migration to 30~AU place firm upper limits on the Kuiper belt mass of just
20 Earth masses \citep{gml04,hm99}, so Jupiter likely received no more than 0.1 Earth masses of material from the Kuiper
belt or beyond.


Even if the proper temperatures are somehow reached, additional problems exist for the clathrate model.  Laboratory
results at very low pressures are sparse and have not shown that clathration rates at the very low nebula pressures are
sufficiently fast to operate on Myr timescales \citep{hgt08}.  And, in order for ice to trap other volatiles as
clathrates, its surface must remain exposed (`microscopic' grains or cracks) for several Myr until the nebula cools
\citep{hgl04}.  This is inconsistent with models showing that grains grow rapidly; planetesimals up to 100~km are
thought to be able to form within 1~Myr even at 30~AU \citep{wei97}.  If for any of these reasons clathration is
inefficient, then Jupiter's core mass could grow much higher than observed due to high O:H ratios required.  Finally,
this existence of such small grains would work to insulate the disk and keep it warm, rather than allow it to cool as
required.  Thus, the solar system's bodies may simply have formed too large, too quickly, to clathrate sufficient
volatiles from the disk atmosphere.

All the models discussed do a reasonable job of fitting Jupiter's composition.  The SCIP models predict a uniform
enhancement of 3$\times$ in all species, fitting the original Galileo data very well.  However, no SCIPs have ever been
identified in the Solar System: comets, for instance, are depleted substantially in N and are of decidedly non-SCIP
composition \citep{igh03}.  The clathrate models result in too much S (by a factor 2$\times$).  All of the models
attempt to fit the roughly $3\pm1\times$ enhancement of the original Galileo values, and do not try to fit the newer
$4\pm2\times$ values; \citet{oe06} dismisses the revised values as being possibly due to systematic errors.  None of the
models fit the high Ar value of $5.4\pm1.1\times$ of GAS07, though the $2.5\pm0.5$ value of Lod08 clearly fits more
easily.  Ar has in fact the \textit{coldest} condensation temperature ($\sim$35~K into clathrates, and 20~K as a solid)
of all the species measured, so its high GAS07 value would be a particular challenge to all the condensation models.

Finally, \citet{lod04} proposed an entirely different model: that Jupiter formed from carbon-rich (rather than ice-rich)
planetesimals.  However, this scenario depends on the assumption that the Galileo probe sampled a typical region of
Jupiter's atmosphere, and most work supports the opposite view that the probe hit an anomalously dry spot.  

\section{Our model: A contaminated molecular cloud}

We propose an entirely different solution to the problem.  Rather than forming Jupiter and the Sun from identical `Solar
nebula' material and invoking condensation or transport within the nebula to modify Jupiter's composition, we propose
that the composition of Jupiter and the Sun differ because of intrinsic temporal and spatial variations in the Solar
nebula composition as the interstellar medium (ISM) is polluted by massive stellar winds and supernovae (SNs).  We
show that this model can explain Jupiter's chemical composition, easily fits into environmental formation scenarios, and
is consistent with other heterogeneities in both our Solar System and distant star clusters.

\begin{figure}
\centerline{\scalebox{0.3}{\includegraphics{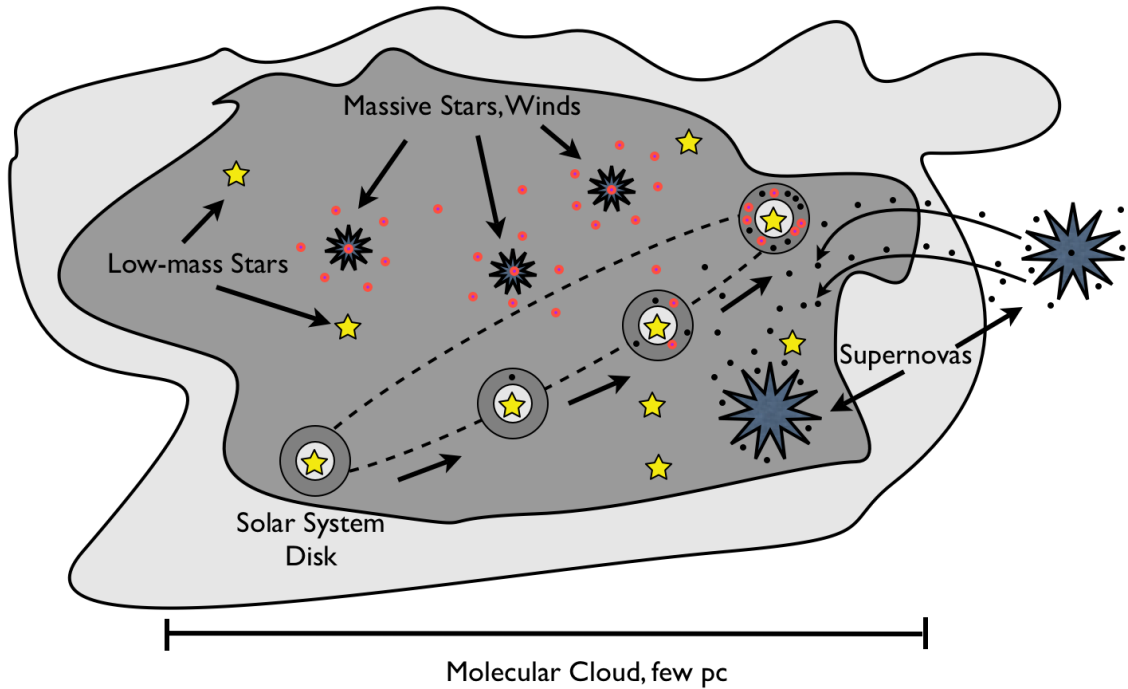}}}
\caption{The proposed `polluted accretion' scenario for Jupiter's atmosphere.  In this model, the Sun and its disk form
in a low-metallicity molecular cloud (left).  The Sun's orbit takes it through other regions (right) of higher
metallicity, polluted by massive stellar winds and supernovae.  Ongoing Bondi-Hoyle accretion from the ISM delivers this
enriched material to the disk, where it is incorporated into Jupiter's atmosphere.}

\label{fig:cartoon} 

\end{figure}

In the scenario that we propose, multiple stages of star formation occurred within a giant molecular cloud
(Fig.~\ref{fig:cartoon}).  The cloud was of average size, 10--20~pc, and within the GMC lay several pc-scale molecular
clouds.  Stars, including the Sun with its disk, began to form in clusters within these clouds.  The Sun's orbit through
the cluster took it on a long path several pc across.  Before any nearby O/B stars turned on, the cluster environment
remained cool and dark for several Myr.  During this time the Sun passed through the ISM and gravitationally swept up
material onto its disk by Bondi-Hoyle accretion \citep{tb08}.

If the Sun formed alone, then the composition of the ISM might be uniform.  But in our proposed scenario,
other clusters nearby formed a few Myr earlier, some with higher-mass stars, and the ISM surrounding these clusters soon
began to be polluted in heavy elements produced by these massive stars.  Nucleosynthesis in these stars enriched their
ejecta by $\sim$100$\times$ or more relative to Solar H.  The Sun's evolving orbit took it through these more
polluted regions of the ISM.  Material was accreted onto both the Sun and its disk, but the disk's large cross-section
and low mass made it easier to pollute than the Sun.  The disk's metallicity slowly increased, and Jupiter's core
and atmosphere formed from the disk.  The disk dispersed within 5--10~Myr, around the same time as the local ISM
dispersed and the cluster spread apart, ejecting the Sun as an unbound field star.  Jupiter's final composition thus
represented largely the same material as the Sun, but with the late accretion of polluted material reflecting in
Jupiter's enriched composition today.

Our proposed model reflects much of the current understanding of star formation in large clusters and molecular clouds
\citep[\eg][]{bal08}.  Stars do not form in isolation, the ISM is not of uniform composition, stars travel on long orbits
through their young clusters, and material from the ISM can be accreted onto young stars and disks in the several
Myr after they form.  All of these are newly appreciated processes that have not been incorporated into existing models
of Solar System formation.  We find Jupiter's metallicity to be one natural consequence of these processes, and we
explore here the issues involved with it.  In the following sections we examine the detailed chemical constraints
on Jupiter's composition from such a model (\S\ref{sect:chemical_constraints}), and the timing and spatial
parameters required (\S\ref{sect:timing_constraints}).  In \S\ref{sect:other_examples} we describe several other cases
where stellar pollution occurs.

\section{Chemical constraints}

\label{sect:chemical_constraints}

In this section we investigate the detailed chemical enrichment from this `polluted accretion' scenario and whether it
can explain Jupiter's current composition.  

We start by assuming that Jupiter's atmosphere is composed of a mixture of three distinct components.  The majority of
the mass is made of Solar material, whose composition is well determined by the measurements of GAS07 and AG89.
The remainder is the small amounts of pollution that come from massive stellar winds and/or supernovae.  
The compositions of these ejecta are distinct `fingerprints' of the stars, determined mostly by their initial mass
and composition.  In general, the winds of these stars are enriched primarily in light elements (C, N, O, Ne), while the
SN ejecta several Myr later contain heavier species (S, Ar, Kr, Xe).  As one example, the demise of a star with an
initial mass of 20 \msol\ can produce 5 \msol\ of O, 0.5 \msol\ of C, 0.2 \msol\ of Si, 0.001 \msol\ of $^{26}$Al, and
10$^{-4}$ \msol\ of $^{60}$Fe in its SN phase \citep[][hereafter WH07]{wh07}.  In our models, we use the grid of winds
and ejecta computed by WH07.  These models span the mass range 12--40~\msol, and consider a
variety of nucleosynthesis rate coefficients and explosion energy parameters, for a total of 66 models.

We denote the elemental abundances in the Solar, wind, and SN components as $n_{\sol}$, $n_{\rm{w}}$, and
$n_{\rm{sn}}$; the Jupiter abundance is $n_{\rm{J}}$.  It is then possible to calculate model Jupiter compositions by
using \begin{equation} n_{\rm{J,i}} = f_{\sol}\, n_{\sol,i}  + f_{\rm{w}}\, n_{\rm{w,i}}  + f_{\rm{sn}}\,
n_{\rm{sn,i}}, \label{eq:linear_combination} \end{equation} where $i$ is the species and $f_{\sol}$ + $f_{\rm{sn}}$ +
$f_{\rm{w}}$ = 1.  We use the present-day solar abundance, but the difference between this and the Sun's primordial
composition is insignificant for our purposes.

Using Eq.~\ref{eq:linear_combination}, we found combinations of solar material and ejecta that would produce
Jupiter's measured composition.  We searched all $66^2 = 4356$ possible combinations of the 66 wind and 66 SN models of
WH07, coupled with the single solar abundance.  For each trial, we computed coefficients that best fit Jupiter.  Our
routine attempted to fit only the well measured stable species (C, N, S, Ar, Kr, Xe), and computed results for both
these and the remaining elements (He, O, Ne, P).

\begin{figure}
\centerline{\scalebox{0.7}{\includegraphics*[75,200][500,350]{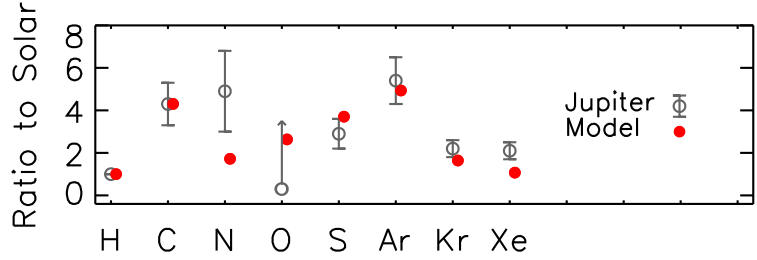}}}
\caption{Elemental abundances of Jupiter from Galileo probe (grey), and fits for our `Model A' case (red).  The
y axis plots the elemental number abundances relative to the Solar abundances, normalized to hydrogen.  The model
consists of a linear combination of 87\% solar composition, 9\% from stellar winds from a 40~\msol\ star,
and 4\% ejecta from an SN of original mass 20~\msol.  O is a lower limit because the Galileo probe entered Jupiter
at a cloud-free location believed to be anomalously dry.  Solar and Jupiter compositions are revised 2007 values
(GAS07, WLA08).} \label{fig:fit_wla08} 

\end{figure}

\begin{figure}
\centerline{\scalebox{0.7}{\includegraphics*[0,200][500,350]{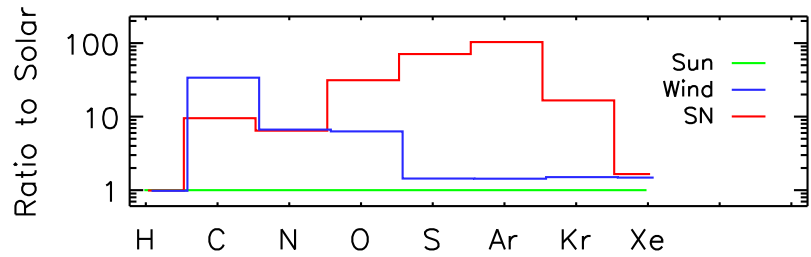}}}
\caption{Sources of individual species in our `Model A' case.  The model is a linear combination of elemental abundances
from the Sun (green curve), stellar winds (blue curve), and an SN (red curve).  Each line is normalized to H at 1.0.  The
stellar winds produce much of the C and N, while the supernova supplies most of the remaining species.} 
\label{fig:fit_wla08_components} 
\end{figure}

Our best fit (`Model~A') is shown in Figures \ref{fig:fit_wla08}--\ref{fig:fit_wla08_components}.  This model finds
Jupiter's composition to be well described by 87\% solar nebula, 9\% stellar winds from a 40~\msol\ star (WH07's model
\texttt{s40a28A}), and 4\% supernova ejecta from a 20~\msol\ star (WH07's \texttt{s20a37n}).  The total contamination is
13\% (\ie, 0.13~\mjup).  The fit is excellent at matching the observed quantities of C, S, Ar, and Kr, and the lower
limit for O.  The largest deviation is for N, where we are slightly below the error bar.  The wind predominantly
supplies C, N, and O while the SN supplies the remaining species.  Both stars have high enough mass (and thus short
enough lifetimes) that they can form and explode within the 10~Myr timeframe of GMCs.  We assume the GAS07 and WLA08
values for the Solar and Jovian composition.  This model is shown here because it is the best fit; many other
combination of different mass SN and wind ejecta provided much worse fits.

\begin{figure}
\centerline{\scalebox{0.7}{\includegraphics*[75,200][500,350]{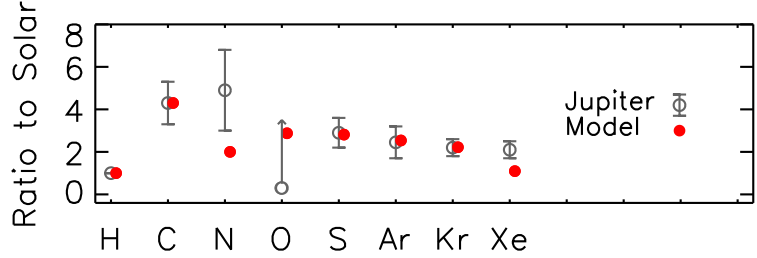}}}
\caption{Elemental abundances of Jupiter from Galileo probe (grey), and fits for our `Model A2' case (red).  
The model consists of a linear combination of 78\% solar composition, 8\% from stellar winds from a 40~\msol\
star, and 14\% ejecta from an SN of original mass 20~\msol.  Abundances are the same as Model~A, except the Lod08 argon
value is used.} \label{fig:fit_lod08} \end{figure}
\begin{figure}
\centerline{\scalebox{0.7}{\includegraphics*[0,200][500,350]{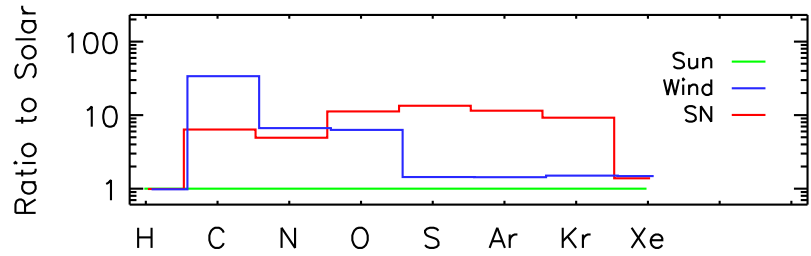}}}
\caption{Sources of individual species in our `Model A2' case.  Same as Figure~\ref{fig:fit_wla08_components}, but
using Lod08 argon value.} 
\label{fig:fit_lod08_components} 
\end{figure}

A variant of this fit ('Model~A2') is shown in Figures \ref{fig:fit_lod08}--\ref{fig:fit_lod08_components}.  In
this case we have used the latest Lod08 value for Ar, instead of that of GAS07.  This model is fit with 78\% solar
nebula, 8\% winds from a 40~\msol\ star (WH07's \texttt{s40a28A}), and 14\% SN ejecta from a 15~\msol\ star (WH07's
\texttt{s15a34c}).  The only difference to the fit is the Ar abundance.  Because lower-mass SNs are less efficient at
heavy-element nucleosynthesis, the fit here requires a substantially larger SN contribution than does Model~A (14\% vs.
4\%).  The 15~\msol\ star has a lifetime of $\sim$11~Myr, putting it on the upper end of individual cloud lifetimes but
within the timescale of large regions like Orion.

\begin{figure}
\centerline{\scalebox{0.7}{\includegraphics*[75,200][500,350]{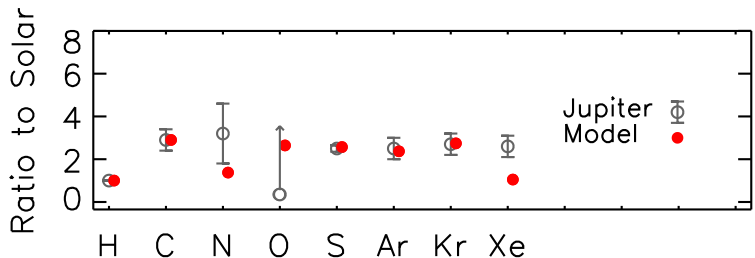}}}
\caption{Our `Model B' fits.  Same as Figure~\ref{fig:fit_wla08}, but assuming 1989 solar composition data (AMN03,
AG89).  The coefficients are Solar (94\%), stellar winds from 40~\msol\ star (4\%), and a supernova from a star of
original mass 25~\msol\ (1.5\%).  The fractional contamination in this model is
6\%, less than half that in Model A.  The low predicted value for Xe might be explained by its particularly
high condensation temperature (see text).   The broad similarity of the results to the `Model A' case shows that our
model is robust against small changes to knowledge of the composition of the Sun and Jupiter.} 

\label{fig:fit_amn03} \end{figure}

\begin{figure}
\centerline{\scalebox{0.7}{\includegraphics*[0,200][500,350]{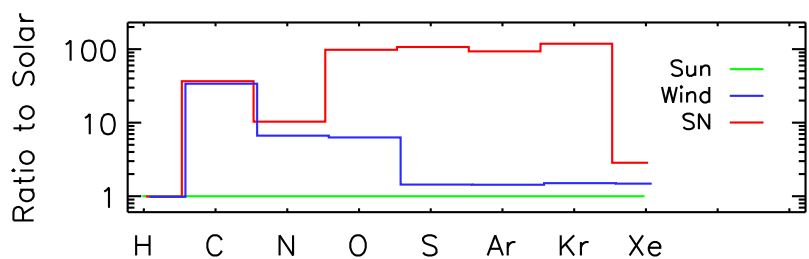}}}
\caption{Individual components of our `Model B' fits.  Same as Figure~\ref{fig:fit_wla08_components}, but using
1989 solar abundances (AMN03, AG89).} 
\label{fig:fit_amn03_components} 
\end{figure}

Finally, a third fit is shown in Figures \ref{fig:fit_amn03}--\ref{fig:fit_amn03_components} (`Model B').  This model
differs in that we have used the abundances of AG89 and AMN03 for $n_{\rm{sol}}$ and $n_{\rm{Jup}}$.   Although the newer
abundances are probably preferred, using the old ones gives a test of the robustness of our fits.  Also, the
nucleosynthetic yields of WH07 start with the AG89 solar abundances, so in a sense this fit is more self-consistent,
even though it is based on slightly older data.  Model B requires about 6\% total contamination, less than half that
required by Model A.  Model B is comprised of 94\% solar, 4\% stellar winds from a 40~\msol\ star (\texttt{s40a28A}),
and 1.5\% SN ejecta from a 25~\msol\ star (\texttt{s25a41d}).

All three of our models fit the data well.  In all cases the vast majority of Jupiter's mass (78\%-95\%) comes from the
Solar nebula, in agreement with standard models.  And in each, a combination of winds and SN -- which one would expect
in a realistic cluster -- works better than any single component by itself.  All three fits are slightly low in N and
Xe; the remaining species are fit very well.

The individual `fingerprints' of the SNs can be seen in Figures~\ref{fig:fit_amn03_components},
\ref{fig:fit_lod08_components}, and \ref{fig:fit_wla08_components}.  In all the models, the lighter species (C, N, O)
are produced in the winds, while the heavier elements come from the SNs and are highly dependent on the SN mass. 
A large difference between the three models is the SN C:N ratio, which is several times higher in Model B than the
others (compare red curves in \ref{fig:fit_lod08_components} and \ref{fig:fit_wla08_components}).  The great deal of carbon
ejected by the Model B SN allows the total contamination in this model to be about half that in Model A.  

Detailed yields from our three models are listed in Table~\ref{table:results}.  This table lists additional species
measured by Galileo, but which are not in equilibrium at the entry site and thus not expected to fit: He, Ne, and P.
Helium and Ne are believed to combine into He-Ne `raindrops' which sink to Jupiter's interior, and cannot be used as a
constraint \citep{rs95}.  Helium itself is produced by the Sun so its primordial Solar abundance cannot be directly
measured.  O was measured at Jupiter but its value is believed to be anomalously low due to the probe's dry entry site.
Encouragingly, our models predict primordial Jovian abundances for O, Ne, and P similar to those of the other species.
The predicted values for O are in the range 2.6--2.9, similar to global O values determined spectroscopically (WLA08).

Several changes could improve the quality of our fits.  First, we have used a simple fitting method, assuming
contamination by only one wind and one supernova.  In realistic star-forming regions such as Orion there are several
dozen stars within a few pc all above 8~$\msol$ that will explode as SN; using multiple stars will increase the ease
of fitting Jupiter's composition.  Second, the SN ejecta models we use are quantized in relatively large mass bins
(5~\msol), and all use certain common assumptions for stellar and explosion parameters.  The SN ejecta yields are very
model dependent; for instance, the models of \citet{yf07} vary in abundance for individual species by 50\% or more
from those of WH07 for stars of similar mass.  The WH07 yields ignore stellar rotation, which may be important
\citep{hmm05}.  Thus, a better understanding of SN yields could help (or hurt) our fits.  Finally, we have not included
physical processes in the ISM or disk such as sedimentation, concentration, or condensation, even though these could be
important.  For instance, of the noble gases our poorest fit is for Xe, which is low by 50\% and at the edge of the 1
$\sigma$ error bar.  However, Xe condenses at a substantially warmer temperature than any of the other noble gases (at
$\sim$55~K into clathrates, it is the easiest to condense), so if our model were to include condensation explicitly, it
would operate in the direction to correct this deficiency.  For now, however, we have intentionally chosen to keep our
models simple to demonstrate that good fits are possible even with a very limited set of parameters.


\section{Stellar Pollution into the ISM}

\label{sect:timing_constraints}

Now that we have shown that Jupiter can be matched chemically, we investigate the requirements on the environment to
support such contamination.

As massive stars in a cluster evolve, they perform nucleosynthesis and create heavy elements that pollute the ISM.
These are given off in stellar winds (during the stellar lifetime) or SNs (at the end-of-life for stars with
M~$>$~8~\msol).  After being injected into the ISM, these highly enriched ejecta are incorporated into the next
generation of stars.  The highest-mass stars are the shortest lived: for instance, the 40~\msol\ stars used in our fits
explode in less than 5~Myr, allowing for a full generation of star formation within the 10~Myr timescales of planet
formation and embedded clusters.

\subsection{Contamination by Stellar Winds}
Stars spend the majority of their lives on the main sequence.  During this stage the stellar winds are weak enough
that they do not pollute the cluster; for instance, the current solar wind is $\mdot_{\sol} \approx 2\times 10^{-14}
\msolyr$ with $v \approx 400\ \kms$.  Red supergiant (RSG) stars, however, have slow massive winds that can easily
pollute the cluster.  RSGs are normal post-main-sequence stars of mass 5--15~\msol\ that have cooled dramatically
after their main-sequence phase (T $>$ 10,000~K to T $<$ 3000~K), and have heavy and slow winds of \mdot $\approx
10^{-5}$--$10^{-4}\ \msolyr$ and $v$ as low as 10~\kms\ -- \textit{ten order of magnitude} more loss than the present-day Sun
\citep{kw93}.  The RSG phase lasts for 5--10\% of the stellar lifetime, or typically a few $10^5$--$10^6\ \rm{yr}$
\citep{ssm92}.  For stars $>25$~\msol, over half the original stellar mass can be lost during the RSG phase
\citep{glm96}.  The wind velocity of 10~\kms\ is only slightly higher than the stellar and gas velocities, so this
material can readily mix with the local ISM.  The abundances $n_{\rm{w}}$ that we consider in
\S~\ref{sect:chemical_constraints} are predominantly from the RSG phase, because stars of less than 20~\msol\ evolve too
slowly ($\gt$ 10~Myr) to enter the RSG phase within typical cluster lifetimes.  Our Model A and B fits use winds from
stars of 40~\msol\ with lifetimes $\sim 5\ \rm{Myr}$, which easily fit the timing constraints.  These high-mass stars
are seen in clusters of $N>$ a few thousand; an example is $\theta^1$~Ori~C in the Orion Trapezium core.  Smaller
8~\msol\ stars are produced in clusters of $N>$ a few hundred. 




\subsection{Contamination by Supernova Ejecta}
After passing through the RSG phase, stars with masses $>8\ \msol$ usually end their lives as SNs.  The SN explosion
gives off 1--5~\msol\ of metal-enriched material at speeds of 2,000--10,000~\kms.  Even at such
high speeds, numerical simulations by \citet{odh07} showed that disks at 1~pc are very resilient to nearby explosions.
However, although the disk will absorb most of the intercepted solids such as \isotope{Fe}{60}, most of the SN's gas
is deflected by the disk.  For a 20~\msol\ explosion at 1~pc, the disk will absorb just $10^{-4}$~\mj\ of enriched gas,
insufficient for the process described here.

But, there are at least two mechanisms by which ejecta may cool and then be accreted onto the disk.  First, as the
ejecta spreads, it mixes with the ISM until it slows and cools, and this cooler ejecta can be accreted more easily.  In
order to slow to 10~\kms, 1~\msol\ of SN ejecta must mix with roughly 1000~\msol\ of ISM (\ie, a 1000:1 mixing ratio).
The SN ejecta can be quite clumpy, with high mixing ratios in some regions and low in others.  Rather than mixing
uniformly, these clumps appear to be slowed as a unit and preserve their density, much like a baseball is slowed in the
air without fully mixing.  Observations of pc-scale ejecta from various SNs show highly clumpy knots on AU scales or
greater maintained for a year after explosion \citep{fhh07,wh94}.  Observations of emission lines even 700 years after the
explosion continue to find evidence for density contrasts of 100--200 in dense metal-rich bubbles within the ejecta
\citep{wbr08, fhm06}.  Additional work shows that some heterogeneity persists on even longer Myr timescales \citep{fhy06,
khf06}.  Thus, diluting to a 1000:1 mixing ratio is needed to slow the material, there will likely be local regions
where the mixing ratio could be 100:1 or lower, roughly the amount of SN pollution required by our `Model B' case.  The
GMC itself is generally robust against disruption by SNs, as evidenced by the Orion region, which has seen dozens of
supernovae in the past 12~Myr \citep{bal08}, yet remains largely a cool GMC.  



The second mechanism involves AU-scale `droplets' of ejecta that can travel from a supernova exploding as a
`champagne fountain,' through the ISM, and land cool and intact in molecular clouds many kpc distant.  This mechanism is
believed to be responsible for observed compositional heterogeneities (`abundance discrepancy factors', or ADFs) of
50\%--10$\times$ in molecular clouds \citep{meg08,str07,tp05,ten96}.  These ADFs are seen in cold, dense,
hydrogen-depleted gaseous regions in the middle of larger molecular clouds, on scales of 10--100~AU.  They are commonly
seen in star-forming clouds including Orion and 30~Doradus.  The droplets have mixing lifetimes of several Myr,
consistent with the timescales needed for our Jupiter model.  In this way, dense concentrations of gaseous metals produced by
the SN are transported to and preserved in the nebula, where they are then passed on to stars and planets that form in
the region \citep{str07}.

\subsection{Accretion of the ISM onto disks}

Once enriched material has mixed with the ISM by either of these two methods, accretion onto the disk is
straightforward.  \citet{tb08} found that the average ISM-to-disk accretion rate for disks in young clusters was
$10^{-8}\ \msolyr$, or $\sim$1~MMSN per Myr.  The total accretion needed in our Model B case is $0.1\ \mjup = 10^{-2}\
\rm{MMSN} = 10^{-4}\ \msol$, or the amount delivered in $\sim$ 10,000 years.  Since accretion may be maintained for 5
Myr, this provides more than sufficient mass delivery.  Due to the disk's large cross-section and its low mass,
accretion affects the disk but makes only negligible effect on the stellar composition \citep{mt09}.


%

\label{sect:other_examples}

\section{Heterogeneity in the Solar System and star-forming regions}

\label{sect:isotopes}

Our model reflects the growing body of evidence that the Solar System did not form from a homogeneous cloud in an
isolated environment, but rather from a heterogeneous nebula where interactions with its environment
played a major role in shaping its evolution.  We briefly discuss here four examples of large-scale
heterogeneity: two in the terrestrial planets, one in star-forming regions, and one on galactic scales.


The terrestrial bodies have been modified by scores of chemical and physical processes since their formation.
Most of these processes act equally on all isotopes of the same element, so isotopic differences are not expected in
samples formed from a well-mixed nebula.  However, isotopic variations \textit{have} indeed been measured
in samples from the Earth, Mars, and numerous asteroids.  Isotopic differences have been measured for species including
Ba, Cr, S, Ti, Zi, Mb, O, and more \citep[][and references therein]{tba07,rj06,dmr02}.  These anomalies are small (up to 1\%
for \isotope{Cr}{54}/\isotope{Cr}{53}, and as small as ppm for some others) but are indisputable.  UV photochemistry has
been invoked to explain the origin of the O variations \citep{ly05}, but the remaining heterogeneities have defied
explanation by known fractionation processes.  Instead, they are consistently thought to be of nucleosynthetic origin,
resulting from the incomplete mixing of ejecta from multiple SNs in the material of the young solar nebula
\citep{rj08,tba07,rj06,dmr02}.



%

In addition to the stable isotopes, variations have been found in short-lived radioactive isotopes, allowing some
constraint on the timing of the Solar System's early heterogeneity.  \citet{knb08} examined the heterogeneity of
\isotope{Al}{26} isotopes within Ca-Al inclusions (CAIs) in carbonaceous chondrites.  They found differences of
$>100\times$ in primordial $^{26}$Al abundances, leading to the conclusion that there are at least two populations of
CAIs: some which were formed in the presence of $^{26}$Al, and some which were not.  This requires either spatial or
temporal variations in the Solar System birth environment.  Their preferred interpretation is that the $^{26}$Al-free
CAIs were formed early (possibly during initial collapse of the Solar System), followed by late injection of $^{26}$Al
from a nearby massive star.  They suggested that this massive star could be the same one that injected $^{60}$Fe several
Myr later after an SN explosion.  We caveat this point with mention that a recent study of CAIs within a single ordinary
chondrite did not reproduce the \isotope{Al}{26} variations \citep{vcl09}; it is possible there are regions that were
heterogeneous and regions that were not.

Orion, the nearest massive star-forming region, shows evidence for `polluted accretion' amongst individual young
stars.  Observations of the metallicity of 29 B, F, and G stars within Orion have found abundance variations up to
4$\times$ between stars of the same age in the same subgroup \citep{csp00,csl98,cl94}.  These variations are spatially
correlated, suggesting that recent supernovae have contaminated distinct regions of the cluster.  The abundance
variations are seen in O and Si (which are produced by massive stars and type II SNs), but not in Fe, C, and N (which
are produced to a far lesser degree, and thus would not be expected to contaminate the stars).  The authors proposed
that ejecta from recent SNs have been accreted onto these stars, causing their observed metallicity.

Relative to the galactic value, the entire Solar System exhibits a system-wide excess of 30\% in $^{18}$O/$^{17}$O, and
the solar value is higher than nearly every molecular cloud or YSO within 10~kpc.  This enhancement has been proposed to
be due to accretion of supernova ejecta and massive stellar winds immediately prior to the proto-Solar nebula's
formation \citep{ygs09,ygs08}.

These four examples describe heterogeneity broadly similar to what we propose for Jupiter.  The contamination at Jupiter
is many times higher than the isotopic anomalies in the terrestrial bodies.  However, the terrestrial anomalies were
formed earlier in the solar system's history, over a shorter period, and in a warmer and better-mixed region.  Jupiter
formed later and in the outer solar system, where chemical pollution can more easily be acquired from the ISM, and mixing
times are far longer.  While the small terrestrial anomalies were likely to have been inherited from incomplete mixing
of the \textit{initial} solar nebula, Jupiter's enhancement could be caused by \textit{post-collapse} contamination of
the disk, from Bondi-Hoyle accretion onto the disk over several Myr and several pc, allowing for far more introduced
pollution across the entire molecular cloud.  The amount of heterogeneity seen at Jupiter \textit{is} comparable or
smaller than that seen in both star-forming regions (ADFs) and young stars themselves.  Because the physical processes
of star formation are universal, if the Sun formed in a dense cluster, then the solar system and Jupiter could have
naturally inherited these heterogeneities.

\section{Discussion}

The model we propose here is a departure from existing models for the solution to Jupiter's metallicity problem.
Although historically most Solar System formation models have assumed a homogeneous `Solar nebula' composition, our
model explicitly assumes the opposite.  We incorporate the fact that the composition of the solar nebula can change
spatially and temporally.  These changes occur during the Solar System's first 5--10~Myr, as the Sun is traveling
through its birth cloud, experiencing the environmental effects of other stars.  The inclusion of this contamination
reflects the latest understandings of the environmental processes affecting star formation.



Our model has three general advantages over the existing amorphous ice and clathrate models that we describe in
\S~\ref{sect:previous}.  First, it relaxes the very strict low-temperature requirements for the formation of Jupiter's
solids.  Second, it allows for Jupiter to have formed at its present location without migration.  Third, it explains the
fact that different elements are enriched by different amounts (\ie, not a uniform 3$\times$).

Our model is not without problems.  Our fits for N are at the edge (or a little beyond) the $1\sigma$ level.  Nitrogen
is poorly measured in the Galileo probe data, but is a major species and an important constraint.  We could increase the
contribution from stellar winds (which supply much of the N), but this would increase the amount of C beyond that
observed.  Low-mass stars of 4--8~\msol\ produce large quantities of N sufficient to solve the problem, but these stars
have lifetimes of 30~Myr or more before they enter the RSG phase.  The other specie our models are all low in, Xe, has
a particularly high condensation temperature (making it easy to condense), and this may in part explain its abundance.

Second, the amount of SN contamination required is on the high end of what can be supplied by nearby SNs using direct
injection of ejecta into a nearby GMC.  Additional work is necessary to understand the mixing and concentration
mechanisms of this ejecta.  However, if the `droplet' model of SN ejecta is correct, then concentrated ejecta can be
slowed and cooled over kpc distances, which would solve this problem.

Most work suggests that Saturn, Uranus, and Neptune accreted their atmospheres from the disk in much the same way
Jupiter did.  The metallicity enrichments of the outer planets exceed Jupiter's -- Uranus and Neptune have 30 times the
Solar C:H ratio, for instance -- suggesting that they were formed at least in part by the traditional chemical
condensation series \citep{lew72}.  The colder nebula and slower evolution make volatile condensation easier at greater
distances.  However, `accreted pollution' of the disk from the ISM would affect these planets as well.  Because of their
different positions in the disk and their different formation times, our model cannot \textit{predict} the enrichment
they may receive from pollution, except that it could be similar to Jupiter's.  (The outer disk, with its larger
cross-section, might be more easily contaminated, but this requires more study.)   More detailed analysis of this awaits
better knowledge of their bulk atmospheric compositions, which are largely unknown today.  Jupiter's oxygen composition
will be probed by Juno in 2017 and will provide a discriminant between many models including our own (which predicts a
global O:H ratio of $\sim$ 2.6$\times$ solar) and that of Lodders~2004 (which assumes $\sim 0.35\times$ solar).  

Jupiter's atmosphere may have formed through a variety of processes, of which polluted accretion could be only one.  For
instance, N is among the most difficult species to condense in the various ice models, as it requires a temperature
$\leq$~30~K.  Nitrogen, however, is easily supplied in the cool, slow stellar winds that we study here.
It is possible that N was supplied by these winds, and other species were delivered in part by icy planetesimals from a
relatively warm ($\geq$ 50~K) disk.

An advantage of the SN model is that it provides a mechanism for matching not only the chemical abundances which we
study here, but also isotopic differences.  It has already been shown that SNs may explain the isotopic differences seen
in the Solar System's rocky bodies (\S~\ref{sect:isotopes}); additional work on the isotopic differences between the Sun
and Jupiter would be a natural extension of the present work.  This path is particularly ripe for future study given the
recent Solar composition results from the Genesis mission \citep[\eg,][]{wbh07,mmh07}.

The general process of accretion from ISM $\rightarrow$ disk $\rightarrow$ planet that we describe here has broader
applications for the formation of extrasolar planets.  A strong observed correlation exists between the metallicity of
the host star and the existence of extrasolar planets \citep[\eg,][]{ufq07}.  We predict that the correlation between
stellar and planet metallicities will depend on the formation environment.  Stars that form in dense regions, in the
presence of massive stars, will have the highest disk metallicities and thus the greatest chance of forming planets.


\section{Acknowledgments} We thank S. Atreya, W. Bottke, F. Ciesla, A. Heger, T. von Hippel, H. Levison, A. Morbidelli,
and M. Wong for useful discussions.  Heger and S. Woosley also kindly provided the ejecta compositional data which we use
here.  HT and JB graciously acknowledge support from NASA Origins grants NNG06GH33G and NNG05GI43G; HT acknowledges
support from NASA Exobiology grant NNG05GN70G, and JB acknowledges support from the University of Colorado Center for
Astrobiology, which is supported by the NASA Astrobiology Institute.

\begin{table}[h]
\centerline{\scalebox{0.45}{\includegraphics{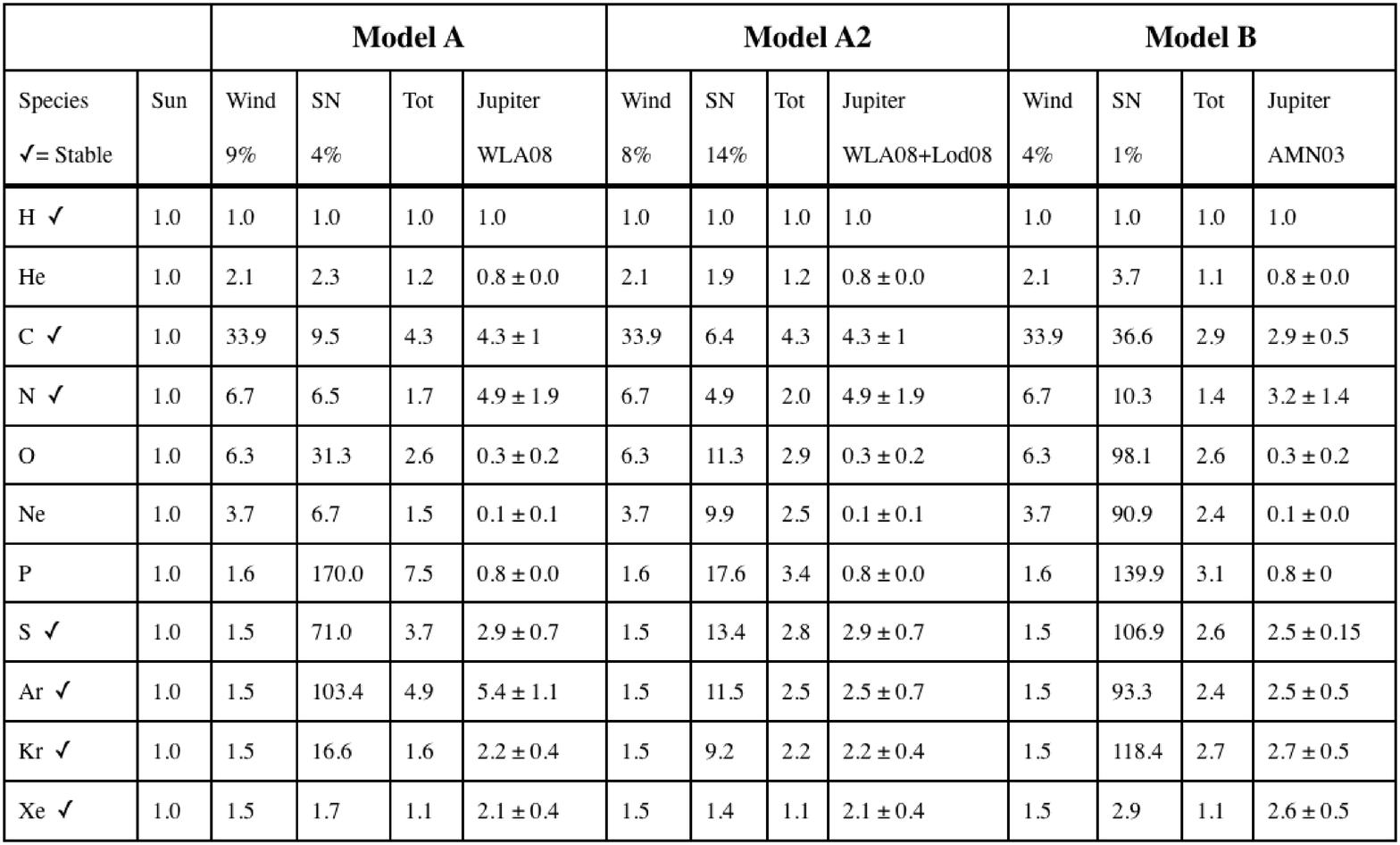}}}
\caption{Details of model results, based on the listed ejecta yields from winds and SN compared with Jupiter.  Table
lists species observed by Galileo; abundances are in number density relative to H, normalized to Solar.  The checked
species are those which are stable at Jupiter and our model attempts to fit; we predict the unchecked species but do not
attempt to fit them.  The three models are based on different measurements for the Jupiter:Solar abundances.}
\label{table:results} \end{table}

\vfill
\eject
\clearpage

\bibliography{papers}

\end{document}

%% file: htmac.tex


%


\def \ApJ	 	{Astrophys. J.}
\def \apj 		{Astrophys. J.}
\def \apjl 		{Astrophys. J. Lett.}
\def \apjs 		{Astrophys. J. Suppl.}
\def \AJ 		{Astron. J.}
\def \aj 		{Astron. J.}
\def \aap 		{Astron. \& Astrophys.}
\def \aaps 		{Astron. \& Astrophys. Suppl.}
\def \apss 		{Astrophys. \& Spac. Sci.}
\def \mnras 		{Month. Not. Royal Astron. Soc.}
\def \nat 		{Nature}
\def \grl 		{Geophys. Res. Let.}
\def \jgr 		{J. Geophys. Res.}
\def \mps 		{Meteorit. Plan. Sci.}
\def \pasp 		{Pub. Astron. Soc. Pac.}
\def \planss 		{Plan. Spac. Sci.}
\def \baas 		{Bull. Amer. Astron. Soc.}
\def \ao		{Appl. Optics}
\def \araa		{Ann. Rev. Astron. Astrophys.}
\def \jqsrt		{J. Quant. Spect. Rad. Trans.}
\def \ssr		{Space Sci. Rev.}


\def \cms		{ cm s$^{-1}$ }			
\providecommand{\kms}   {\ensuremath{\rm{km\ s^{-1}}}}
\def \gcm		{ g cm$^{-2}$ }			
\def \per		{ $^{-1}$ }			
\def \pertwo		{ $^{-2}$ }			
\providecommand{\sol}	{\ensuremath{{\odot}}}		
\providecommand{\msol}	{\ensuremath{M_{\sol}}}		
\providecommand{\lsol}	{\ensuremath{L_{\sol}}}		
\providecommand{\esol}	{\ensuremath{E_{\sol}}}		
\providecommand{\mjup}	{\ensuremath{M_{\rm{J}}}}	
\providecommand{\mj}	{\ensuremath{M_{\rm{J}}}}	
\def \water		{H$_2$O}			
\providecommand{\tonec}		{\ensuremath {$\theta^1$ Ori C} }		
\newcommand \isotope[2] {\ensuremath{{}^{#2}\rm{#1}}}   

\def \paa		{Pa$\alpha$}			
\def \ha		{H$\alpha$}
\def \Ha		{H$\alpha$}
\def \hii		{H{\small II}}
\def \sii		{S{\small II}}
\def \oiii		{O{\small III}}
\def \arcmin		{'}
\def \eqtext#1		{\hspace{1in} \hbox{#1}}	
\providecommand{\micron}{\ensuremath{\mu\rm{m}}}
\providecommand{\tpp}   {\ensuremath{\tau \varpi_0 P}}	
\providecommand{\rgi}	{\ensuremath{\rm{r}_{gI}}}
\providecommand{\rgii}	{\ensuremath{\rm{r}_{gII}}}
\def \etal      	\hbox{ \it et al.} 		
\def \etals		\hbox{{\it et al.}'s}		
\def \vs		\hbox{vs.}			
\providecommand{\degrees}{\ensuremath{{}^\circ}}
\providecommand{\mearth} {\ensuremath{M_{\oplus}}}	
\def \b			{$\bullet\ $}
\def \Beta		{\beta}				
\def \yes        	{$\surd\ $}     		
\def \idlplot#1		{\centerline{\scalebox{0.9}{\includegraphics{#1}}}} 
\def \idlplotps#1	{\centerline{\scalebox{0.9}{\includegraphics[70,350][574,710]{#1}}}} 
\def \ie		{{\it i.e.\/}}
\def \eg		{{\it e.g.\/}}
\def \subsimt#1{{\lower 2pt\hbox{$\scriptstyle #1$}\atop
     \raise 1pt\hbox{$\scriptstyle \sim$}}}
\def \gtrsim    	{\subsimt >}			
\def \lessim    	{\subsimt <}
\def \lesssim    	{\subsimt <}

\providecommand{\gt} 	{\ensuremath{>}}
\providecommand{\lt} 	{\ensuremath{<}}


\newcommand{\rj}	{\ensuremath{R_{\hbox{J}}}}


\newcommand{\rbh}	{\ensuremath{R_{\rm{B}}}}
\newcommand{\rtidal}	{\ensuremath{R_{\rm{T}}}}
\newcommand{\racc}	{\ensuremath{R_{\rm{Acc}}}}
\newcommand{\dmdt}	{\ensuremath{\dot{M}}}
\newcommand{\dmdtbh}	{\ensuremath{\dot{M}_{\rm{B}}}}
\newcommand{\mdotbh}	{\dmdtbh}
\newcommand{\dmbh}	{\ensuremath{{\Delta M}_{\rm{B}}}}
\newcommand{\dmdtacc}	{\ensuremath{\dot{M}_{\rm{acc}}}}
\newcommand{\dmdtstar}	{\ensuremath{\dot{M}_{*}}}
\newcommand{\msolyr}	{\ensuremath{M_{\odot}\ \rm{yr^{-1}}}}
\newcommand{\msolmyr}	{\ensuremath{M_{\odot}\ \rm{Myr^{-1}}}}
\newcommand{\mmsnmyr}	{\ensuremath{\rm{MMSN}\ \rm{Myr^{-1}}}}
\newcommand{\mdot}	{\ensuremath{\dot{M}}}


\newcommand{\promille}{ 
  \relax\ifmmode\promillezeichen
        \else\leavevmode\(\mathsurround=0pt\promillezeichen\)\fi}
\newcommand{\promillezeichen}{%
  \kern-.05em%
  \raise.5ex\hbox{\the\scriptfont0 0}%
  \kern-.15em/\kern-.15em%
  \lower.25ex\hbox{\the\scriptfont0 00}}


\newcommand{\qej}	{\ensuremath{q_{ej}}}


\newcommand{\ltsim}	{\ensuremath{\lesssim}}
\newcommand{\gtsim}	{\ensuremath{\gtrsim}}


\def \sk		{\vskip 0.1 in}

\def \doublespace 	{\baselineskip = 24 pt}
\def \singlespace 	{\baselineskip = 12 pt}
\def \halfspace 	{\baselineskip = 18 pt}

\font\bigrm = 		cmr10 scaled \magstep 1
\font\bigbigrm = 	cmr10 scaled \magstep 2
\font\halfbigrm = 	cmr10 scaled \magstephalf
\font\bigbf = 		cmb10 scaled \magstep 1
\font\bigbigbf = 	cmb10 scaled \magstep 2
\font\halfbigbf = 	cmb10 scaled \magstephalf

\def \in		{\leftskip = 0.0 in}
\def \ini	 	{\leftskip = 0.2 in}		
\def \inii	 	{\leftskip = 0.4 in}
\def \iniii	 	{\leftskip = 0.6 in}
\def \iniiii	 	{\leftskip = 0.8 in}

\def \hang 		{\parindent = -0.15 in \leftskip = 0.15 in}	
\def \nohang		{\parindent = 0 in \leftskip = 0 in}

\def \figstart		{\leftskip = 0.5 in \rightskip = 0.5 in}	
\def \figend		{\leftskip = 0 in \rightskip = 0 in}	

\def \instart		{\leftskip = 0.2 in \rightskip = 0.2 in}	
\def \inend		{\leftskip = 0 in \rightskip = 0 in}

